\documentclass[12pt,preprint]{aastex}

\newcommand{\bdv}[1]{\mbox{\boldmath$#1$}}

\def\au{{\rm AU}} 
\def\kms{{\rm km}\,{\rm s}^{-1}}
\def\kpc{{\rm kpc}}
\def\pc{{\rm pc}}
\def\max{{\rm max}}
\def\min{{\rm min}}
\def\rel{{\rm rel}}
\def\e{{\rm E}}
\def\bpi{{\bdv\pi}}

\begin{document}
\title{LSST's DC Bias Against Planets and Galactic-Plane Science}

\author{Andrew Gould}
\affil{Department of Astronomy, Ohio State University,
140 W.\ 18th Ave., Columbus, OH 43210, USA; 
gould@astronomy.ohio-state.edu}

\begin{abstract}

An LSST-like survey of the Galactic plane (deep images every 3-4 days)
could probe the Galactic distribution of planets by two distinct
methods: gravitational microlensing of planets beyond the snow line
and transits by planets very close to their hosts.  The survey would
identify over 250 disk-lens/disk-source microlensing events per year
that peak at $r<19$, including 10\% reaching the high magnification
$A>100$ that makes them especially sensitive to planets.  Intensive
followup of these events would be required to find planets, similar to
what is done presently for Galactic bulge microlensing.  The same data
would enable a wealth of other science, including detection of
isolated black holes, systematic study of brown-dwarf binaries, a
pre-explosion lightcurve of the next Galactic supernova, pre-explosion
lightcurves of stellar mergers, early nova lightcurves, proper
motions of many more stars than can be reached by GAIA, and probably
much more.  As usual, the most exciting discoveries from probing the
huge parameter space encompassed by Galactic-plane stellar populations
might well be serendipitous.  Unfortunately, the LSST collaboration
plans to exclude the first and fourth quadrants of the Galactic plane
from their ``synoptic'' observations because the DC image that
resulted from repeated observations would be limited by crowding.
I demonstrate that the majority of
this science can be recovered by employing well-developed
image subtraction analysis methods, and that the cost to other
(high Galactic latitude) science would be negligible.

\end{abstract}


\section{{Introduction}
\label{sec:intro}}

The past two decades have seen a tremendous growth of precision optical
photometric surveys.  The first of these were microlensing surveys
\citep{pac86,pac91,griest91}
carried out by the MACHO \citep{macho}, EROS \citep{eros}, and OGLE \citep{ogle}
collaborations,
which monitored tens of square degrees at several hundred (now thousand) epochs.
These have also yielded an immense range of scientific results that are 
unrelated
to microlensing, from transiting planets \citep{udalski02a}, to proper motion
catalogs \citep{sumi04}, to pre-explosion lightcurves of stellar
mergers \citep{tylenda11} and of other outbursts \citep{wagner12}.
The Sloan Digital Sky Survey (SDSS, \citealt{sdss,sdss2}) 
then mapped $\sim 10^4\,\rm deg^2$ in five bands at a single epoch,
supplemented by a synoptic survey that covered hundreds of square degrees
at several dozen epochs.  Influenced largely by SDSS's spectacular
success (over 5000 publications to date), the \citet{nwnh} gave top
ground-based priority to a new survey that would combine the principal 
features of these two.  Like SDSS, it would cover a large fraction of the
sky, and like microlensing surveys, it would do so at $\sim 10^3$ epochs.
Strangely, however, there has been little discussion of the microlensing
potential of an all-sky synoptic survey.  Investigation of this potential
leads immediately to the conclusion that many other applications
are being overlooked, perhaps at great scientific cost.

Among the properties one would like to know about extrasolar planets
is their distribution in the Galaxy.  This question is particularly
difficult to address because most planet-search techniques depend
on subtle changes in light from the host and so are restricted to
planets orbiting bright (hence generally nearby) stars.  Thus, the
overwhelming majority of the roughly 4000 planets and strong planetary
candidates detected to date are within a kpc or so of the Sun.

The microlensing technique is an important exception to this rule.
Planetary systems give rise to microlensing signals when they 
act as ``lenses'', deflecting the light of unrelated, more distant
``source'' stars.   Hence, detection does not depend in any way on
the light from the host, nor even the existence of a host.  The
principal challenge for the microlensing technique is that the
source and lens must be aligned exquisitely on the sky, typically
$<1\,$mas, to generate a microlensing signal.  Even for the densest
star fields toward the Galactic bulge, the chance that any given
source is microlensed is only about $10^{-6}$.  Moreover, only a small
fraction of microlensed stars yield planetary signatures.  Hence,
all planetary microlensing surveys have been carried out toward
the Galactic bulge and, ipso facto, all microlensing planets have
been discovered in this direction.

Because the source stars in these surveys
are near the Galactic center, current microlensing
surveys are already sensitive, in principle, to planetary systems at
all Galactocentric radii $R$ in the range $0\la R\la R_0\simeq 8\,$kpc.
However, one of the prices that microlensing must pay for being indifferent
to host light is that generally the hosts are not detected, so that 
neither their masses nor their distances are known.  Fortunately, in the case
of lenses with planetary signatures, one almost always 
sees ``finite source effects''
when the source passes over or near a ``caustic'' induced by the planet.
In these cases, one can directly measure $\rho=\theta_*/\theta_\e$ from
the lightcurve, where $\theta_*$ is the source radius,
$\theta_\e$ is the angular Einstein radius, 
\begin{equation}
\theta_\e \equiv \sqrt{\kappa M\pi_\rel};
\qquad
\kappa\equiv {4 G\over c^2 {\rm AU}} = 8.1 {{\rm mas}\over M_\odot},
\label{eqn:thetae}
\end{equation}
$M$ is the lens mass, and 
$\pi_\rel$ is the lens-source relative parallax.  Since $\theta_*$ can be
determined from the dereddened color and magnitude of the source
\citep{yoo04}, this means that a combination of the host mass and distance
is almost always measured.

Nevertheless, it is generally quite difficult to disentangle $M$ and 
$\pi_\rel$,
and for this reason the distribution of microlens planets along the
line of sight toward the Galactic bulge remains unknown.

An alternate approach would be a microlensing survey of the Galactic
disk.  I argue here that such a survey would be much more sensitive
to the distribution of planets as a function of $R$ than surveys toward
the bulge, in part because it is easier to estimate the distribution
of planet distances and in part because a larger range of $R$ is surveyed.

Of course, such a survey would be much more challenging than current
microlensing planet searches.  Thousands (rather than dozens) of 
square degrees would have to be monitored because the event rate per
unit area is much lower.  There is no foreseeable way to monitor such
a huge area with the roughly hourly cadence required to find planets.
Instead, there would have to be one wide-field survey that monitored the target
fields every few days from which one would identify the promising
microlensing candidates, and then these candidates would have to be 
monitored much more frequently from a network of narrow-angle telescopes
\citep{gouldloeb}.

Fortunately, the Large Synoptic Survey Telescope (LSST) is well adapted to the
first component of this search.  It plans to observe 15,000 square degrees
every 3--4 nights (covering a total of more than 20,000 deg$^2$)
over the course of 10 years to flux levels that are
sufficient to find all microlensing events that could plausibly be followed
up.  The network of narrow-angle telescopes required for the second
component, high-cadence followup observations, is already under construction.
Hence, such a survey is quite feasible.

Unfortunately, the current plans of the LSST consortium are to
exclude from normal-cadence observations the majority of Galactic disk 
that is accessible to LSST, including all of the disk within $90^\circ$
of the Galactic center where the overwhelming majority of ``disk-disk''
(disk-lens/disk-source) microlensing events occur.

Up until now, the prospects for very wide field microlensing surveys
have only been considered to fairly bright magnitudes.  In particular,
\citet{han08} calculated event rates over the whole sky with magnitude
limits of $V=12$, 14, 16, and 18.  However, due to 
``magnification bias'', which favors the detection of highly magnified,
faint sources, together with the much higher sensitivity to planets of
high-magnification events, most sensitivity to planets comes from
substantially fainter sources than these, which happen to be
momentarily magnified to near the faintest of these limits.

Here, I investigate this potential.  I then briefly review a subset of the other
vast scientific opportunities that such a survey would enable.  Finally,
I discuss the DC bias that is leading the LSST consortium to ignore these
possibilities.

\section{{Microlensing Event Rate in the Galactic Disk}
\label{sec:rate}}

 From the standpoint of finding planets, what constitutes a 
``microlensing event'' is determined by what can plausibly be
monitored in follow-up observations.\footnote{I note that the problem
of time-variable photometry in crowded fields has been solved by the
invention of difference imaging \citep{alard98}, which is routinely
implemented in fields that are much more crowded than those discussed here
by all microlensing teams, who achieve a systematics floor of a few mmag
when reasonable ($\sim 1^{\prime\prime}$) seeing obtains.}
For present purposes, I define
this as events for which the source 1) enters the Einstein ring ($u_0<1$)
2) reaches magnitude $r_{\rm peak}<19$ at the peak of the event, and
3) has a baseline magnitude $r_{\rm base}<26$.  The third requirement
is needed because it would be difficult or impossible to recognize
fainter sources that satisfy (2) in time for intensive monitoring
over the peak.  For simplicity, I consider only disk-disk lensing,
and also consider only main-sequence sources and lenses.  Of course,
giant-star sources are also microlensed, but these also strongly degrade the
microlensing signal of the main-sequence sources that are superposed within
$1^{\prime\prime}$.  For similar reasons, giant-star lenses are rare and
virtually unobservable.  

I adopt a simple double-exponential profile
for both the stars and the dust.  The scale heights of these are 300 pc
and 130 pc, respectively, while the scale lengths are both 2500 pc.
For the local normalization of the dust, I adopt $dA_r/d\ell = 0.6\,
{\rm mag}\,\kpc^{-1}$.
For the stars, I adopt a local luminosity function of
$\phi=(0.34, 0.48, 1.087, 1.96, 2.51, 3.15, 3.02, 3.89, 3.78, 6.97,$
$10.90, 
15.14, 8.86, 5.48, 2.50, 2.50, 2.50, 2.50)\times 10^{-3}\,{\rm pc}^{-3}$
for $M_V=1,\ldots,18$, with corresponding masses
$M=(2.0,1.7,1.4,1.2,1.0,0.9,0.83,0.65,0.51,0.39,0.32,0.25,0.20,0.17,0.13,$
\break
 $0.10,
0.09,0.08)\,M_\odot$.  I assume that the Galaxy has a flat rotation curve
characterized by $v_{\rm rot}=235\,\kms$, with local dispersions in the
radial, tangential, and vertical directions of $(35,28,18)\,\kms$, 
that the Sun is moving with respect to the Local Standard of Rest at
$(10,12,7)\,\kms$ and that it lies 15 pc above the Galactic plane.  

I evaluate the event rate for each source star $i$ 
at distance $D_{S,i}$ by a sum over lenses at $D_{L,j}$,
\begin{equation}
\Gamma(D_S,l,b) = \sum_j 2\theta_{\e,ij} D_{l,j}^2\mu_{\rel,ij}
\rightarrow
2\phi_M\int_0^{D_{S,i}} d D_L D_L^2 {\rho(D_L,l,b)\over \rho_0}
\biggl({\au\over D_L} - {\au\over D_S}\biggr)^{1/2}
\langle{\mu_\rel(D_S,D_L,l,b)}\rangle
\label{eqn:gammadlb}
\end{equation}
where $\theta_{\e,ij}$ is the Einstein radius of the lens (Equation 
(\ref{eqn:thetae})), $\mu_{\rel,ij}$ is the source-lens relative proper 
motion, $\rho(D_L,l,b)/\rho_0$ is stellar density at the lens location relative
to the local one, $\langle{\mu_\rel(D_S,D_L,l,b)}\rangle$ is the
mean magnitude of the lens-source proper motion at the specified coordinates,
and
\begin{equation}
\phi_M \equiv \sum_k \phi_k\sqrt{\kappa M_k} \simeq 0.131\,{\rm mas^{1/2}\,pc^{-3}}
\label{eqn:phim}
\end{equation}

I then evaluate the total event rate per square degree for each line of
sight $(l,b)$ by
\begin{equation}
\Gamma(l,b) = 
\biggl({\rm deg\over radian}\biggr)^2
\sum_k \phi_k \int_0^\infty d D_s D_s^2 u_\max(r[M_{r,k},D_S,l,b])\Gamma(D_s,l,b),
\label{eqn:gammalb}
\end{equation}
where $r(M_r,D_S,l,b) = M_r + A_r + 5\log(D_S/10\,\pc)$, $A_r$ is calculated
by integrating through the above-assumed dust profile from the Sun to the
source, and 
\begin{equation}
u_\max(r) = \min\Biggl(1,\sqrt{2\biggl({A_\min\over \sqrt{A_\min^2 - 1}}-1\biggr)}
\Biggr)\Theta(26-r);
\quad
A_\min = \max\biggl(\sqrt{9\over 5},10^{0.4(r-19)}\biggr)
\label{eqn:umax}
\end{equation}
is obtained by inverting the \citet{einstein36} point-lens magnification:
$A(u)= (u^2 + 2)(u^4 + 4u^2)^{-1/2}$.

Figure \ref{fig:rhombus} shows the resulting rates in 
events $\rm yr^{-1}\,deg^{-2}$ for the region within $20.5^\circ$ of
the Galactic plane.  The total rate in the area shown is
$\Gamma = 557\,\rm yr^{-1}$ of which 86\% is inside the dashed-black
rhombus, which contains just $2076\,\rm deg^2$, i.e., 14\% of the total 
area shown.  Note from the lower panel of Figure \ref{fig:rhombus} that
the dashed-black rhombus also contains the great majority of events
for which the lenses are more than a few kpc from the Sun.

\section{{Planet Searches}
\label{sec:planet}}

Of course, not all of these $\Gamma = 557\,\rm yr^{-1}$ could be 
detected.  First, most regions of the Galactic disk come too close
to the Sun to be observed all year.  Second, any given telescope will
be restricted to observing those portions of the sky that do not remain
too close to (or below) the horizon at that site.  Nevertheless,
a telescope at a southern site could observe most or all of the
rhombus shown in Figure~\ref{fig:rhombus} for more than half the year,
and therefore could detect of order 250 events per year, or more.
This is of order 15\% of the rate of event detection by the OGLE 
collaboration \citep{ogle} toward the Galactic bulge in its
OGLE-IV phase\footnote{http://ogle.astrouw.edu.pl/ogle4/ews/ews.html}.
And it is roughly 35\% of the current rate for the MOA collaboration
\citep{ob03235} in 
MOA-II\footnote{https://it019909.massey.ac.nz/moa/alert2013/alert.html}
phase or the previous detection rate of OGLE-III.  In the period 2007-2009
when OGLE-III and MOA-II were in operation, microlensing planets
were being discovered at a rate of roughly 3 per year despite the
fact that less than 5\% of the discovered events were being aggressively
monitored by follow-up collaborations, such as 
PLANET \citep{ob05390}, $\mu$FUN \citep{ob05169}, 
RoboNet \citep{mb10073}, and MiNDSTEp \citep{ob08510}. 
Hence, it is plausible that aggressive
monitoring of these 250 events could yield a dozen planet detections
per year.

I note that there is a strong magnification bias induced by selecting on
peak flux, so that a disproportionate share of the events will be at
high magnification even relative to the OGLE-III survey.  See
Figure~\ref{fig:umin}.  Half the events have $A_\max>6$ and thus
significantly enhanced sensitivity to planets \citep{gouldloeb}, while
10\% have $A_\max>100$, which implies greatly enhanced sensitivity
\citep{griest98,gould10}.

However, given that the magnified flux threshold required to achieve
this event rate is $r_{\rm peak}<19$, whereas the densely monitored
Galactic bulge events typically had $I_{\rm peak} \la 15.5$, it is clear
that Galactic-plane followup observations would require much more
telescope resources than earlier bulge searches.  This requirement
is not as exacting as it may sound: many earlier planet detections
relied heavily on amateur class telescopes (e.g., \citealt{ob120026}).
What is needed now is
a dedicated network of 1m class telescopes.  Fortunately, such a 
world-wide network is currently under construction by the 
Las Cumbres Observatory Global Telescope Network (LCOGT)
\citep{tsapras09}, with planet discovery being an important component
of its science program.  

\section{{LSST: An Ideal Disk Microlens Survey}
\label{sec:lsst}}

Therefore, the most difficult aspect of organizing such a planet search
is to conduct the underlying survey that will identify the microlensing
events for high-cadence followup.  This survey must satisfy four
key requirements: (1) observe a large fraction of the Galactic plane
(2) from the southern hemisphere (3) every few days (4) to very deep
magnitudes.  These characteristics can mostly be inferred by examination
of Figures~\ref{fig:rhombus} and \ref{fig:umin}.

From Figure~\ref{fig:rhombus}, it is clear that most of
the survey sensitivity comes from the dashed-black rhombus centered
on the $(l,b)=(0,0)$, whose area is $\sim 2000\,\rm deg^2$.  Since its
center has a declination of $\delta\sim -30^\circ$, 
both of the first two criteria 
are necessary.  Much of the potential planet sensitivity comes from
high-magnification events.  For example, \citet{gould10} found 6 planets
in a sample of 13 events that were well-monitored over peaks of $A_\max>200$.
Figure~\ref{fig:umin} shows that there are 35 such events per (full) year,
with about half these occurring in the rhombus at times when they are
observable.  But if such an $A_\max=200$ event has a typical timescale of
$t_\e=50\,$days and peaks at my adopted $r_{\rm peak}=19$ threshold, then
3 days prior to peak, it will be $A=50/3\sim17$ and so have a magnitude
$r = r_{\rm peak} + 2.5\log(200/17)= 21.7$.  To recognize such an incipient
high-magnification event therefore requires observations every 3--4 days
with good enough (few percent) photometry to both identify the most recent
point as ``interesting'' and to recover the previous lightcurve in order
to predict its future behavior.  This justifies the final two criteria,
``(3) every few days (4) to very deep magnitudes''.
Note that the photometry does not have to be good enough to unambiguously
predict future behavior.  As with current followup photometry carried out 
by $\mu$FUN, it is enough to identify candidate high-magnification events that
can be checked on subsequent nights by a few observations, to then determine
whether they are suitable for intensive followup over peak.

Figure~\ref{fig:rhombus} shows the approximate boundaries of the 
high cadence zone of the proposed
LSST survey in Galactic coordinates.  These are derived from
the equatorial-coordinate diagram from Figure 18 of \citet{ivezic13},
which is an updated version of Figure 4.4 of \citet{lsst}.  The
white contours indicate the limits imposed by geography, i.e., the
northern and southern declinations at which the air mass begins
to seriously degrade the quality of observations.  The dashed-black
rhombus at the center is the region that is excluded for reasons
discussed in Section~\ref{sec:DC}.  Table~\ref{tab:eventrate}
summarizes the integrated event rate inside and outside this rhombus,
both for the entire area shown in Figure~\ref{fig:rhombus} and for the
region robustly accessible to LSST.

\begin{table}
\caption{\label{tab:eventrate} \sc Event Rates by Zone}
\vskip 1em
\begin{tabular}{@{\extracolsep{0pt}}llrr}
\hline
\hline
Subset   & Rhombus & Event Rate & Area  \\ \hline
(Fig.~1) &         & ($\rm yr^{-1}$) & ($\rm deg^2$) \\ \hline
\hline
All      & Inside  &      477 &  2076  \\ \hline
All      &Outside  &       80 & 12411  \\ \hline
LSST     & Inside  &      431 &  1685  \\ \hline
LSST     &Outside  &       47 &  5639  \\ \hline
\end{tabular}
\end{table}

\section{{Advantages of Disk-Disk Lensing}
\label{sec:diskdisk}}

While Galactic disk fields have many fewer microlensing events than
bulge fields despite much larger area, they do possess several significant
advantages, which combine to make mass measurements easier.  Recall
from Section~\ref{sec:intro} that $\theta_\e$ is routinely measurable
for planetary events.  This means that the lens mass $M$ can be determined
provided that the microlens parallax $\pi_\e$ can be measured
\citep{gould92,gould04},
\begin{equation}
M = {\theta_\e\over \kappa \pi_\e},
\quad
\pi_\rel = \theta_\e\pi_\e.
\label{eqn:massdis}
\end{equation}
The microlens parallax is a vector $\bpi_\e$ whose magnitude is the
size of Earth's orbit relative to the Einstein radius projected onto
the observer plane $\pi_\e = \au/\tilde r_\e$ and whose direction
is that of the lens-source relative proper motion.

Microlens parallax is usually measured from distortions in the lightcurve
due to the accelerated motion of the observer (on Earth).   Hence,
if the event is very short, Earth's motion can be approximated
as being uniform during the event, so the effect is negligible.
For disk-disk lensing, the observer, lens, and source all share,
to some extent, the motion of the disk, which tends to make the
event last longer than in disk-bulge lensing. 

Second, just by chance,
the Galactic bulge lies very near the ecliptic.  In the limit that
the source lies exactly on the ecliptic, the direction of Earth's
acceleration remains constant.  Since the component of $\bpi_\e$ that
is parallel to this acceleration $\pi_{\e,\parallel}$
is third-order in time \citep{gmb94}
while the perpendicular component $\pi_{\e,\perp}$
is fourth order \citep{smith03,gould04}, virtually all of the
uncertainty is concentrated in one component for bulge sources,
making it exceptionally difficult to measure the amplitude of $\bpi_\e$,
which goes into Equation (\ref{eqn:massdis}).  By contrast, much
of the Galactic disk lies well away from the ecliptic.

Third, for disk-disk lensing, the direction of lens-source relative
proper motion is expected to be approximately aligned with the Galactic plane,
both because dispersions in the radial and rotation directions are
larger than in the vertical direction and because the bulk relative
motions of the local standards of rest of the observer, lens, and, source
lie almost exactly in the plane.  This implies that even for the cases
that only one component of $\bpi_\e$ is measured, there is significant
prior information on the direction of $\bpi_\e$ to deproject its
amplitude, at least statistically.

Finally, in strong contrast to disk-bulge lensing, the {\it source direction}
itself provides important statistical information of the Galactocentric
distance of the lens in disk-disk lensing.  That is, just from the
ratios of planet detections to total microlensing events monitored
along different lines of sight, one already learns something about the 
Galactic distribution of planets.

\section{{Other Synoptic Disk Science}
\label{sec:science}}

While microlensing planet searches are the immediate focus the present
work, there are a range of other applications of the same
many-epoch survey of the Galactic disk.  These all basically stem from
the fact that the majority of Galactic stars lie within the dashed-black
rhombus of Figure~\ref{fig:rhombus}.

\subsection{{Other Microlensing}
\label{sec:micro}}

First, there are other microlensing applications.  Even without the
follow-up observations discussed in this work, the survey itself would
yield the event rate and optical depth over many lines of sight, which
in turn probe the compact-matter distribution of the disk as a function
of Galactocentric radius.  The EROS collaboration carried out such
a study over four non-bulge lines of sight \citep{derue01,rahal09}.
See Figure~1 of \citet{rahal09}.  However, an order-of-magnitude larger
survey is now feasible.  Such an investigation would be complementary
to GAIA kinematic measurements, which will be
similar to the Hipparcos-based studies of 
\citet{creze98} and \citet{holmberg04}, but on a much larger scale.
That is, while the kinematic measurements are sensitive to total
mass (stars, brown dwarfs, gas, dark matter), the  microlensing measurements
are only sensitive to compact objects.  In addition, the two surveys 
may be sensitive to non-standard gravity (e.g.\ \citealt{milgrom83})
in different ways.

The same microlensing data would probe the distribution of isolated black 
holes in the Galactic disk, which cannot be investigated in any other
way.  Because black-hole events are typically very long, they will
generally yield microlens parallaxes and so the parameter combination
$\pi_\e^2=\pi_\rel/\kappa M$.  Such measurements do not unambiguously
identify black holes because small $\pi_\e$ can in principle be produced
by either large $M$ or small $\pi_\rel$.  However, the black holes will
also have large $\theta_\e^2 = \kappa M\pi_\rel$, and this is potentially
detectable through astrometric measurements \citep{my95,hnp95,walker95},
either from the survey itself (see below) or high-resolution followup
observations.

In addition, the survey would yield an immense wealth of data on
the low-mass binaries, including close brown dwarf binaries 
\citep{choi13} over
vast regions of the Galaxy, which cannot be probed in any other
way.

\subsection{{Transiting Planets}
\label{sec:transit}}

Second, the same survey would probe the Galactic distribution of
planets by a completely independent method: transits.  With roughly
$N\sim 800$ epochs, the survey could identify close-in planets
characterized by
\begin{equation}
N\,{R_*\over \pi a}\,{r_p^4\over R_*^4}\sqrt{1-b^2}
\sigma^{-2} >\Delta\chi^2_{\rm thresh}
\label{eqn:transit}
\end{equation}
where $R_*$ is the radius of the star, $r_p$ is the radius of the planet,
$a$ is the semi-major axis of the orbit (assumed circular), $b$ is
the normalized impact parameter, $\sigma$ is the fractional error
of  the measurements, and $\Delta\chi^2_{\rm thresh}$ is the minimum
$\Delta\chi^2$ required for detection.  

For reference, the number of
images is similar to that collected by the OGLE-III project toward two
fields (in Carina and the Galactic center) totaling about $2.2\,\rm deg^2$
\citep{udalski02a,udalski02b},
while the integrations would be several magnitudes deeper and the sky coverage
more than 1000 times larger.
Since the OGLE-III survey found five planets,
these figures alone demonstrate the planet finding potential of such
a survey.

Of course, before making a direct comparison with OGLE, one must take account
of the fact that the $\sim 800$ observations are spread out over 
$T\sim 10\,$yr rather than a few months.  This implies many more ways
to fold the data to search for transits, and hence a higher probability
that purely statistical noise will masquerade as transits.  However,
the following explicit calculation shows that this is not the limiting
factor.

At fixed period $P$ and transit duration $t$, there are $P/t$ independent
locations to search for the transit.  If the period is changed by $\delta P$,
then the offset between the first and last transit will be displaced in time by
$\delta P(T/P)$.  Setting this equal to $t$ yields the period change at
which the set of folds are independent: $\delta P = Pt/T$.  At each $P$
a small (typically factor 5) range of $t$ must be searched, but this
is basically accounted for just by using the minimum $t$ searched.
Hence, to probe periods $P<P_0$  requires $P_0 T/t^2$ independent
searches.  If we demand a false detection
probability $f=10^{-6}$ (which is conservative,
since the transit frequency is about 1000 times higher than this), then
\begin{equation}
\Delta\chi^2_{\rm thresh,stat} \sim 2\ln{P_0 T\over t^2 f}\sim 63
\label{eqn:transit2}
\end{equation}
where I have adopted $T=10\,$yr, $P_0=5\,$day, $t=0.5\,$hr.
For the OGLE survey, A.\ Udalski searched for transits down to
$\Delta\chi^2=81$.  \citet{gould06} conducted double-blind tests
and found that this search was complete to $\Delta\chi^2=121$.
A calculation of the purely statistical limit using 
Equation~(\ref{eqn:transit2}) yields $\Delta\chi^2_{\rm thresh,stat}\sim 56$
for OGLE.  Thus the OGLE survey was not fundamentally limited by
statistical noise, and it is therefore plausible that a future study
could do at least as well.

Now, it must be emphasized that the five OGLE detections required aggressive
followup of about 100 candidates, an effort that would be difficult to
duplicate on a 1000-fold larger scale.  However, by restricting the search
to planets that are smaller than Jupiter (and thus 
also smaller than late M dwarfs and brown dwarfs) and also by focusing on closer
planets, which are both rarer and have higher signal-to-noise ratio
according to Equation~(\ref{eqn:transit}), one could develop a tractable
search program. This would provide an additional valuable probe to the
Galactic distribution of planets at very different host-planet separations
compared to the microlensing search, albeit at closer distances from the Sun.

To make quantitative estimates of this potential, I make the following
assumptions.  First, I adopt the planet frequencies of \citet{howard12},
derived from {\it Kepler} data.  Second, I adopt the single-epoch
photometric precision from \citet{ivezic12}, except that
I adopt a photometric error floor of $2\times 10^{-3}$ rather than the
value of $5\times 10^{-3}$ shown in their Figure 21.  This is because
difference photometry can be done much more accurately than absolute
photometry.  For example the OGLE collaboration routinely achieves a
systematics limit of $4\times 10^{-3}$ using their 1.3m telescope,
while \citet{hartman09} showed that  $1\times 10^{-3}$ can be achieved
for individual sources when data from larger telescopes are carefully 
analyzed.  See their Figure 3.  Finally, I adopt $\Delta\chi^2=121$
which was achieved in the OGLE transit survey \citep{gould06}.
For ``sub-Jupiters'' ($4<r_p/R_\oplus<8$), I find a total of 6000
planet detections within the LSST zone of Figure~\ref{fig:rhombus},
of which 1800 lie in the dashed-black
rhombus. These have a mean distance from the
Sun of about 1 kpc, almost independent of direction.  Of course,
one might argue that since less than one-third of these planets lie
in the rhombus, not much is lost by ignoring it.  However, from the
standpoint of learning about the Galactic distribution of planets,
it is critically important to look both inside and outside the solar
circle.  I also find a total of 200 ``sub-Neptunes'' ($2<r_p/R_\oplus<4$),
with mean distances of $340\,$pc.

\subsection{{Explosions}
\label{sec:explode}}

Another unique application of a deep synoptic survey of the Galactic disk
would be to provide a pre-supernova lightcurve of the next Galactic supernova.
\citet{szczygiel12} are systematically obtaining such lightcurves for about
25 external galaxies with a cadence of a few times per year.  Given
the fact that a supernova in our own Galaxy will be subject to 
extremely detailed study at multiple wavelengths, as well as neutrinos
and perhaps gravitational waves, much more detailed pre-supernova lightcurves
would be of immense value.  This would be true whether the next supernova takes
place during the first such high-cadence photometric 
survey or many decades later.

Similarly, such a survey would obtain pre-outburst lightcurves of a variety
of explosive phenomena, including nova and stellar mergers.  For example,
\citet{tylenda11} measured the inspiral of the eclipsing-binary
progenitor of a merger explosion based on OGLE data.  
\citet{tylenda13} found, unfortunately, that the pre-explosion lightcurve
of another stellar merger began too late to capture the eclipsing phase,
so that the inspiral rate could not be measured.  This points to the
importance of long time baselines over very wide areas.
And the MOA collaboration recognized $P=2\,\rm hr$ periodic variations
in the rising lightcurve of MOA-2012-BLG-320 
(D.\ Bennett 2012, private communication) , a week after it was
triggered at $I\sim 18.5$ as a ``microlensing candidate'', but
five days before it was spectroscopically classified as a nova \citep{wagner12}.
Of course, a 2-hr period could not be recognized in data having a 3-day
cadence, but the rising lightcurve could trigger the required followup.

\subsection{{Proper Motions}
\label{sec:pm}}

\citet{ivezic12} estimate that LSST will attain proper-motion precision
of $\sigma_\mu\sim 200\,\mu \rm as\,yr^{-1}$ for $16<r<21$ and parallax
precision of $\sigma_\pi\sim 500\,\mu \rm as$ (see their Figure~21).
These measurements will not be competitive with GAIA in the magnitude
range that it covers, $V\la 20$, but with modest reddening near the
plane, $V-r\sim 1$ even for G stars.  Hence LSST could obtain 
proper motions for these numerous tracers out to a factor 2.5
farther than GAIA. Note that at, e.g., 8 kpc, $\sigma_\mu$ corresponds
to just $8\,\kms$.  Hence LSST could probe the Galactic potential by
measuring vertical dispersions and find streaming motions in regions
that are either inaccessible to or poorly probed by GAIA.  This same
astrometric precision would be adequate to measure the astrometric
microlensing signatures of black holes mentioned above in favorable cases,
although precision followup would be required in others.

\subsection{{Serendipity}
\label{sec:serendipity}}

There may be many other specific applications of such a data set as well,
but I close this section with a few remarks on serendipitous discovery.
``Serendipitous discovery'' is usually (and correctly) 
justified by past precedent in proposals for surveys
that probe new parameter space.
By its very nature,
the only thing that can be ``predicted'' about serendipity is that
the more parameter space that is probed, the more likely it is that
something interesting will ``turn up''.  A large synoptic survey can
make serendipitous discoveries either outside or inside the Galaxy.
The former potential is basically maximized by viewing as large
a fraction of the low-extinction sky as possible. The latter
is maximized by observing as many stars as possible, which basically
means observing the first and fourth quadrants of the Galactic plane,
which contain the great majority of Galactic stars.  There is some
tension between these two goals because the plane is heavily extincted
and is also ``contaminated'' by stars.  However, the gain that comes
from avoiding the plane is a minor increase in extragalactic cadence,
whereas the loss to Galactic serendipity that comes from avoiding
the plane is catastrophic.

\section{{DC Response to a Dynamic Question}
\label{sec:DC}}

LSST was the highest ground-based priority specified by the \citet{nwnh},
who entitled their report ``{\it New Worlds, New Horizons}'', thereby
placing equal emphasis on extrasolar planets and cosmology as
leading components of the next decade of astrophysical research.
And indeed, their highest priority for space-based research,
a Wide-Field Infrared Space Telescope (WFIRST) actually does give
leading emphasis to these two areas.  However, neither the word
``extrasolar'' nor ``planet'' even appears in \citet{lsst}.
As I have shown in this paper, this is not because the proposed telescope
would be incapable of finding or characterizing extrasolar planets: it
could contribute greatly, primarily via microlensing but also via transits.

\begin{figure}
\plotone{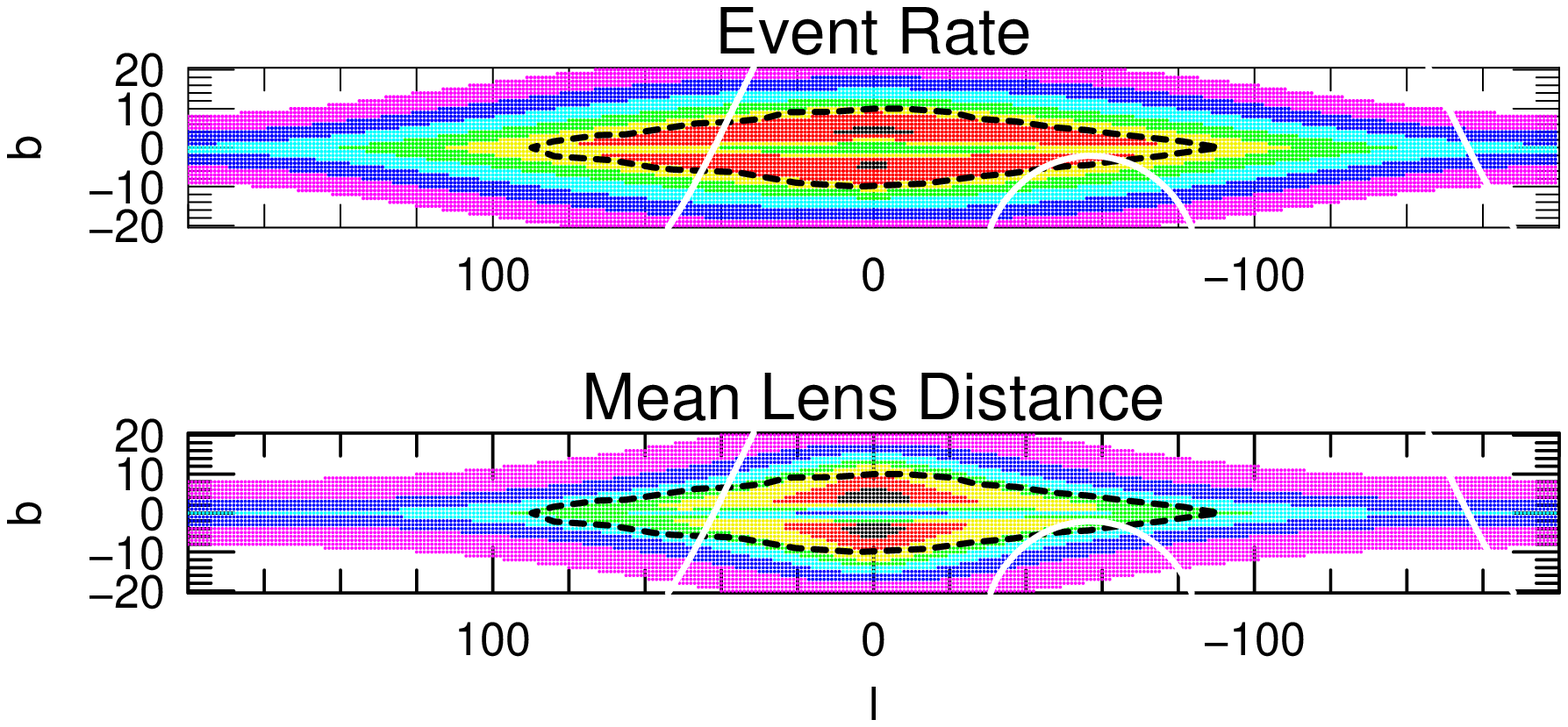}
\caption{\label{fig:rhombus}
Upper panel: Event rate for disk-disk microlensing within
$\pm20.5^\circ$ of the Galactic plane, for events peaking at $r<19$ according
to assumptions described in the text. Zones in black, red, yellow, green,
cyan, blue, magenta
signify rates above $10^{(0,-0.5,-1,-1.5,-2,-2.5,-3)} 
\rm yr^{-1}\,deg^{-2}$.  The regions interior to the white semi-circle
and exterior to the white lines are excluded by LSST because they are
too far south and north, respectively.  The region interior to the
dashed black rhombus is excluded from high cadence observations.  
For the region accessible to LSST, the total microlensing rate is 
$47\,\rm yr^{-1}$ outside the rhombus and
$431\,\rm yr^{-1}$ inside the rhombus.
Lower Panel: Mean lens distance for events shown in upper panel.
Zones in black, red, yellow, green,
cyan, blue, magenta signify mean distances above 5, 4, 3, 2.5, 2, 1.5, 1 kpc.
}
\end{figure}

\begin{figure}
\plotone{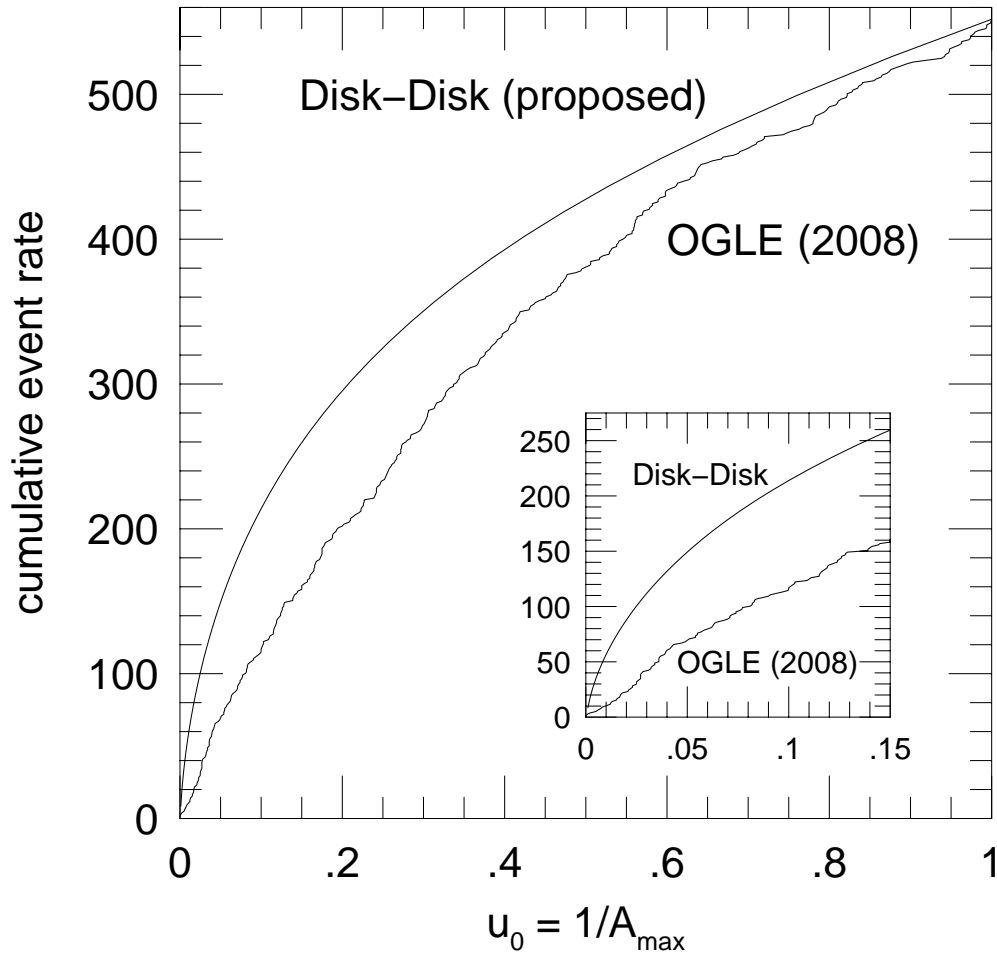}
\caption{\label{fig:umin}
Cumulative distribution of inverse magnifications of proposed disk-disk
microlensing survey compared to actual magnification distribution of
OGLE-III events from 2008 (adapted from \citealt{cohen10}) scaled to the
same total.  Half the events have $A_\max>6$ and thus significantly
enhanced sensitivity to planets \citep{gouldloeb}, while 10\% have $A_\max>100$,
which implies greatly enhanced sensitivity \citep{griest98,gould10}.
}
\end{figure}

The science described in this paper could be achieved if LSST
monitored the regions of the sky available to it (given its
southern location) with approximately uniform coverage.  This
is the region defined by the white contours in Figure~\ref{fig:rhombus}.
In this case, it would cover this region, including a large fraction 
of the dashed-black rhombus, roughly 800 times over 10 years \citep{lsst}.

Unfortunately, LSST plans to avoid the region of the dashed-black
rhombus in Figure~\ref{fig:rhombus}.  The reasoning behind this
decision is not easy to trace.  For example, in their 133 page document
describing the experiment, \citet{lsst} do not mention any analysis
leading to the exclusion zone that appears in their Figure 4.4.

To try to understand the reason for avoiding the plane, 
I interviewed the LSST Project Scientist.
He informed me that there was no point in repeatedly
imaging the Galactic plane because the ``DC image'' (i.e., the co-add of
successive images) would quickly run out of new identifiable sources
due to crowding.  That is, the ``synoptic'' second initial of LSST was
really intended only for extragalactic supernovae, asteroids, and
proper motions of halo stars, but not for the majority of stars in the
Galaxy.  

Close examination of Figure 18 of \citet{ivezic13} shows that no
substantive purpose is served by excluding the ``blue stripe'' in that figure
(dashed-black rhombus in Figure~\ref{fig:rhombus}).  The regions
just to west of the center of the ``blue stripe'' are colored 
``deep brown'' indicating that they are slated for all the ``extra''
observations freed up by the non-observation of the 
``blue stripe''.\footnote{in Figure~\ref{fig:rhombus}, 
this brown region lies just north of the dashed-black rhombus, 
and is centered at
$(l,b)\sim (0,+15)$.}
But this region with extra observations
is not of especially high interest.  From an extragalactic
standpoint, it is much less useful than high-latitude regions because
of higher foreground extinction as well as higher stellar ``contamination''.
And from a Galactic standpoint, it is far less interesting to double
the observations of this region than it would be to observe the
Galactic plane.  

The potential for exoplanetary science, stellar science, and
Galactic-structure science in the plane of the Milky Way is immense
and could easily be achieved at essentially no cost to other LSST science.


\acknowledgments

I thank Zeljko Ivezi\'c for valuable insights about LSST. I thank
Scott Gaudi and Juna Kollmeier for critical readings of the manuscript.
This work was supported by NSF grant AST 1103471.

\end{document}